\documentstyle[preprint,aps,psfig]{revtex}

\input{psfig}
\def\be{\begin{equation}}
\def\ee{\end{equation}}
\def\bea{\begin{eqnarray}}
\def\eea{\end{eqnarray}}
\begin{document}

\title{The Origin of Transverse Flow at the SPS}

\author{M.~Bleicher, C. Spieles, C. Ernst, L. Gerland, S. Soff, 
H.~St\"ocker, W.~Greiner}

\address{Institut f\"ur
Theoretische Physik,  J.~W.~Goethe-Universit\"at,\\
D-60054 Frankfurt am Main, Germany}

\author{S. A. Bass\thanks{Feodor Lynen Fellow of the
        Alexander v. Humboldt Foundation}}

\address{Department of Physics, Duke University\\
        Durham, N.C. 27708-0305, USA}


\maketitle
\begin{abstract}
We study the transverse expansion in central Pb+Pb collisions at the
CERN SPS. Strong collective motion of hadrons can be created. This flow is
mainly due to meson baryon rescattering. It allows to study the 
angular distribution of intermediate mass meson baryon interactions.
\end{abstract}

\newpage
%\vspace*{.5cm}

An important question raised in the recent year is: Is there collective
transverse expansion and flow at midrapidity in relativistic heavy ion
collisions? 
This phenomenon has long been predicted\cite{smg74,st80} as a result of
strong stopping and compression wave or shock wave formation with
subsequent hydrodynamical sideward momentum transfer.
Flow of baryons and mesons has been experimentally observed at the LBL/BEVALAC
and at GSI/SIS in the 1~AGeV regime\cite{kampert,gut}, as well as at the
BNL/AGS\cite{e877,mattiello}.

The question now has been raised whether the strong blue shift of the
inclusive hadron spectra for high multiplicity events observed last year at the SPS
Pb(160~AGeV)+Pb system\cite{NA44,NA49,WA98} can be theoretically explained by
a microscopic calculation. Here we report that the observed strong increase
of $p_{t}$ with $m_{\rm hadron}$ can be due to the large fraction of
secondary meson baryon collisions.

A heavy ion collision can be divided into three phases: Incoming
nucleons interact with each other, they produce secondary particles and
gain transverse momentum. In the course of the rescattering stage, the 
system may reach thermal (or even chemical) equilibrium, while collective
transverse expansion develops. Finally the system dilutes, strong 
interactions cease - the system freezes out.

The studies presented in this letter are performed within the UrQMD
model \cite{urqmd}, a microscopic hadronic transport approach based on the covariant
propagation of mesonic and baryonic degrees of freedom. It allows for 
rescattering and the formation and decay of resonances and strings. 

Let us briefly discuss how meson baryon reactions are modeled within
UrQMD. In the low energy region, i.e. $\sqrt s \stackrel{\textstyle <}{\sim} 2$~GeV,
the inelastic cross section is dominated by s-channel resonance
formation. It is calculated from detailed balance according to the sum
of the possible final state baryon resonances (e.g. $\pi^-+p\rightarrow
\Delta^{0}\mbox{ or }N^*$):
\begin{eqnarray}
\sigma^{\rm MB}_{\rm total}(\sqrt{s}) &=& \sum\limits_{R=\Delta,N^*}
       \langle j_B, m_B, j_{M}, m_{M} | J_R, M_R \rangle \,
        \frac{2 I_R +1}{(2 I_B +1) (2 I_{M} +1 )}  \\
&&\times        \frac{\pi}{p^2_{\rm CMS}}\,
        \frac{\Gamma_{R \rightarrow MB} \Gamma_{\rm tot}}
             {(M_R - \sqrt{s})^2 + \frac{\Gamma_{\rm tot}^2}{4}}\quad,
\end{eqnarray}
which depends on the total decay width $\Gamma_{\rm tot}$, the partial
decay width $\Gamma_{R \rightarrow MB}$, the spins of the particles $I$
and on the c.m. energy $\sqrt{s}$.
According to its lifetime this resonance decays 
{\em isotropically} in its local restframe into (one or more) mesons and a
baryon.

In the ultra high energy limit t-channel processes dominate the total 
cross section, which can be calculated from Regge theory\cite{donnachie92}:
\be
\sigma_{\rm total}=Xs^\epsilon+Ys^{-\eta} \quad.
\ee
Here the first term describes the pomeron exchange, while the second
one is due to $\rho$, $\omega$, $f$ and $a$ meson exchange, while $s$ denotes
the squared center of mass energy. The parameters are fixed by experimental data
when available ($\pi+N$, $K+N$, \dots). For completely unknown cross
sections we employ additional rescaling factors from the additive quark
model\cite{goul83} under the assumption of a 40\%
reduced s-quark cross section (compared to u, d):
\begin{eqnarray*}
\sigma_{\rm AQM} &=& 40\left(\frac{2}{3}\right)^{m_1+m_2}
\left(1-0.4\frac{s_1}{3-m_1}\right)
\left(1-0.4\frac{s_2}{3-m_2}\right)
[{\rm mb}]\quad, \\
\sigma_{\rm MB_{unkown}}(\sqrt s)&=&\sigma_{\pi N}(\sqrt s)
\frac{\sigma_{\rm MB_{AQM}}}{\sigma_{\rm \pi N_{AQM}}}
\end{eqnarray*}
with $m_i =1 (0)$ for particle $i$ being a meson (baryon) and $s_i$
being the number of \mbox{(anti-)}strange quarks in particle $i$.
One or both particles are now excited to longitudinal color flux tubes.
Due to the extremely high relative momenta of the incident hadrons, the
fragmentation of the strings leads to a strongly forward-backward peaked
distribution of secondaries.

Heavy ion collisions at the AGS and, even more so, at the SPS exhibit an
intricate scenario: Here, the typical MB collision energies are in the
range of 2-10~GeV, thus cannot be assigned to one of the above discussed regimes.
The necessity to bridge consistently from
one picture of meson baryon interactions to the other becomes apparent if we look
at the following reactions where a low momentum pion hits a nucleon  
\begin{eqnarray}
\pi + N(938)&\rightarrow&\Delta \rightarrow \pi + N(938)\quad  \mbox{with
 an {\em isotropic} angular distribution,}\\
\pi + N(2000)&\rightarrow&\mbox{String} \rightarrow \pi + N(2000)\quad  \mbox{with
  a {\em longitudinal} angular distribution.}
\end{eqnarray}

To investigate the influence of the modelling of meson baryon scattering
on the collective transverse flow, the UrQMD model was used in two different modes: 
\begin{enumerate}
\item A complete downward extrapolation of the high energy longitudinal
      color flux picture to $\sqrt s = 2$~GeV (refered to as
      forward-backward peaked (f-b) variant) .
\item An upward extrapolation of the resonance behaviour, i.e. an 
      isotropic angular distribution of all outgoing particles in meson baryon reactions
      (indicated as (iso)).
\end{enumerate}
These changes affect only MB collisions above the resonance region
($\sqrt s > 2$~GeV);
all other channels, i.e. meson meson and baryon baryon reactions, remain
unchanged. 

\begin{figure}[t]
\vskip 0mm
%\vspace{0.5cm}
\centerline{\psfig{figure=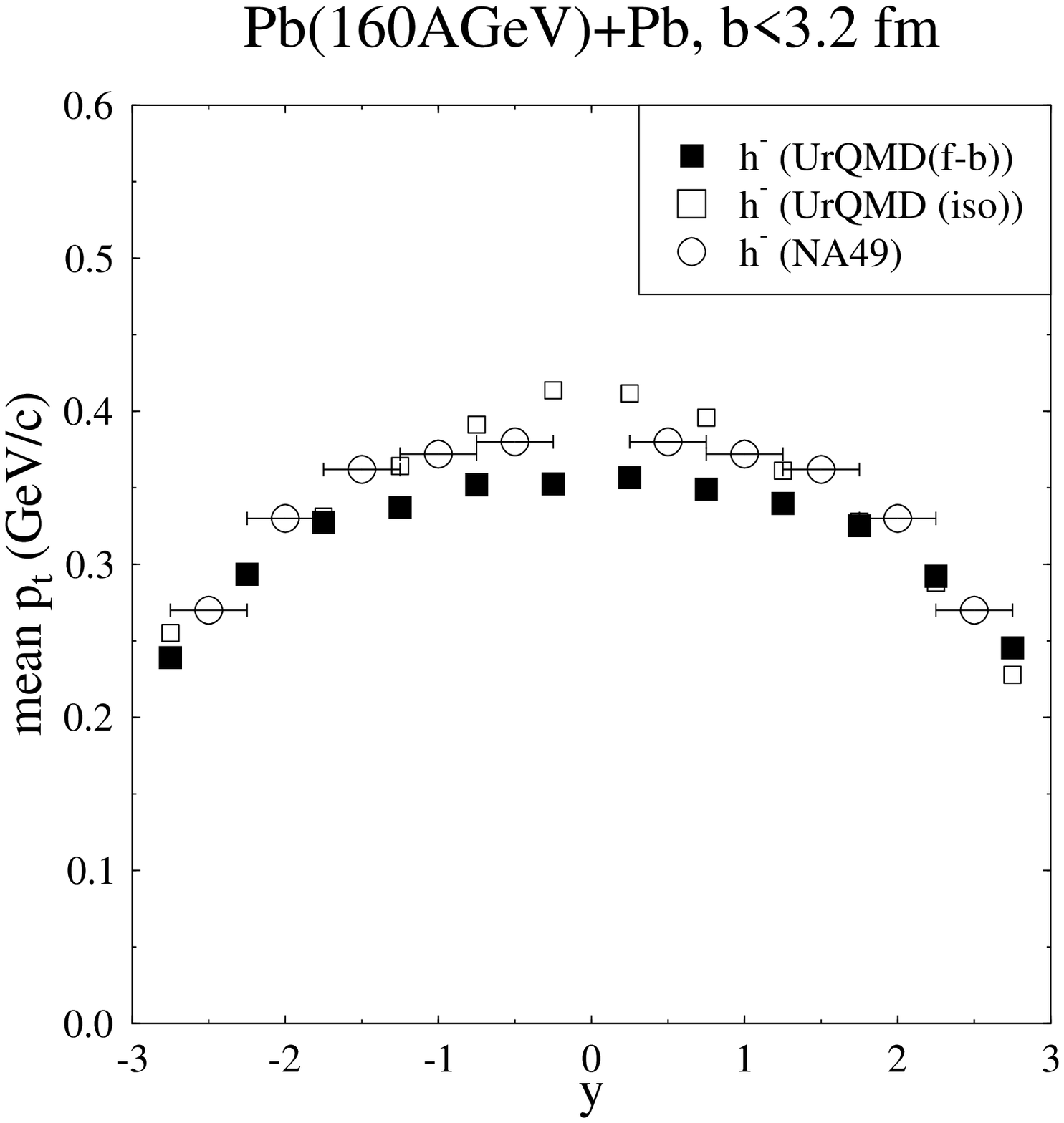,width=10cm}}
\vskip 0mm
\vspace{-.5cm}
\caption{Average transverse momentum of $h^-=\pi^-+K^-+\overline{p}$ in central Pb(160GeV)+Pb
reactions at the SPS.
\label{hminus}}
\end{figure}

Let us start by comparing both scenarios to recent 
preliminary experimental data.
In Fig. \ref{hminus} the mean transverse momentum of negatively charged
particles \mbox{($h^-=\pi^-+K^-+\overline{p}$)} as a function of rapidity in
Pb+Pb, $b<3.2$~fm at 160 AGeV is depicted. Open circles are
NA49 data \cite{NA49}, open squares are calculated with the resonance
extrapolation (iso) and full squares denote the high energy
prescription (f-b). Overall good agreement for both ansatzes with the data 
on the 10\% level is found from projectile to target  rapidity. 

\begin{figure}[b]
\vskip 0mm
\vspace{-1.5cm}
\centerline{\psfig{figure=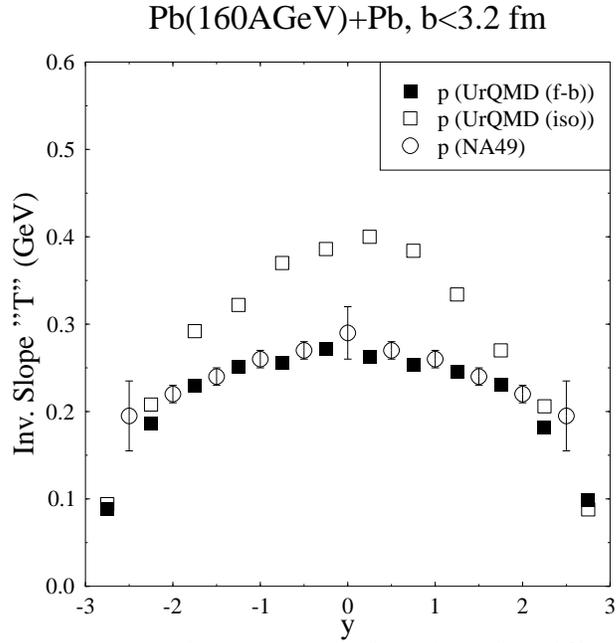,width=10cm}}
\vskip 2mm
\vspace{-1.0cm}
\caption{Apparent temperature of protons as a function of rapidity 
in central Pb(160GeV)+Pb reactions at the SPS.
\label{proton}}
\end{figure}

Let us turn now to the transverse momentum spectra of baryons. 
Fig. \ref{proton} shows the ''apparent temperature''
of protons, ''$T$'', in Pb(160AGeV)+Pb for different rapidities in comparison with preliminary
NA49 data (open circles) \cite{NA49}. 
The inverse slope ''$T$'' is obtained from a fit to the transverse mass
$m_T$ spectrum using 
\be
\frac{1}{m_T}\frac{dN}{dm_T}\propto {\rm e}^{-m_T/T} \quad.
\ee
In both scenarios, the ''temperature''
of the protons increases towards midrapidity. The maximum ''temperatures'' 
are 270 MeV (f-b, full squares) and 395 MeV (iso, open squares). Thus 
the isotropic meson baryon scattering prescription predicts a
temperature at $y_{\rm CM}$ exceeding the preliminary data by 30\%. 

\begin{figure}[t]
\vskip 0mm
%\vspace{0.5cm}
\centerline{\psfig{figure=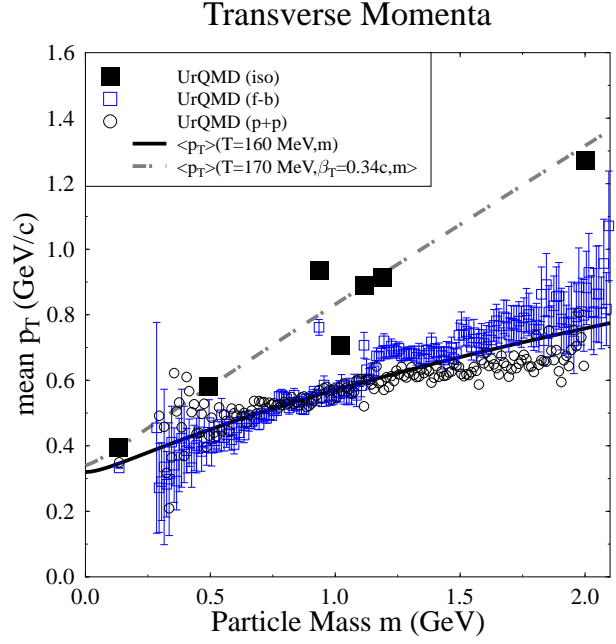,width=10cm}}
\vskip 2mm
%\vspace{0.4cm}
\caption{Mean $p_t$ at midrapidity ($|y|<0.5$) as a function of particle mass
in central Pb(160GeV)+Pb reactions (compared to p(160GeV)+p).
\label{mass}}
\end{figure}

To gain further insight into the creation of transverse flow, 
Fig. \ref{mass} discusses the mean transverse momenta of particles 
for different mass bins at $|y|<0.5$ in central Pb+Pb collisions 
at the SPS. Circles show the p(160 AGeV)+p events, open squares indicate 
the (f-b) Pb+Pb events and the full black squares show the MB-(iso)tropic
model. In addition, we fit the resulting mass spectra with a simplified
2 parameter 
fireball plus flow model. The full black line is the mean transverse
momentum $<p_T(m,T_0,\beta_T)>$ for 
different masses $m$ calculated from a thermal distribution with
temperature $T_0=160$~MeV (to fit the $\pi$) without any additional 
flow ($\beta_T=0$). The dashed 
line shows the $<p_T(m,T_0,\beta_T)>$ from an expanding thermal 
source ($T_0=170$~MeV) with an additional transverse flow velocity 
of $\beta_T=0.34c$. This yields in the saddle point approximation a
rough value for the ''{\em apparent} temperature'' $T$\cite{lee}: 
\be
T\approx T_0+m\beta_T^2 \quad.
\ee
The mean $p_T$ of all hadrons, from proton-proton reactions as well as from
the high energy nucleus-nucleus (f-b) scenario show no
significant difference to a non-expanding thermal source.
However, additional {\em baryon} flow is visible due to further
baryon baryon collisions. This  results in a bump in the mean $p_T$ around
$m\approx 1.1$~GeV. In contrast, the isotropic (MB) model produces a hadronic source
which expands more strongly in the transverse direction. A significant transverse
collective flow velocity of
$0.34c$ is observed in this calculation. Due to its very small cross section in 
nuclear matter the $\Phi$-meson aquires less transverse flow compared to other hadrons.
Note that the error bars are only
statistical. These are given only for 
the (f-b) Pb+Pb reaction, in the (MB iso) and p+p case they are about the same.

\begin{figure}[b]
\vskip 0mm
\vspace{-0.5cm}
\centerline{\psfig{figure=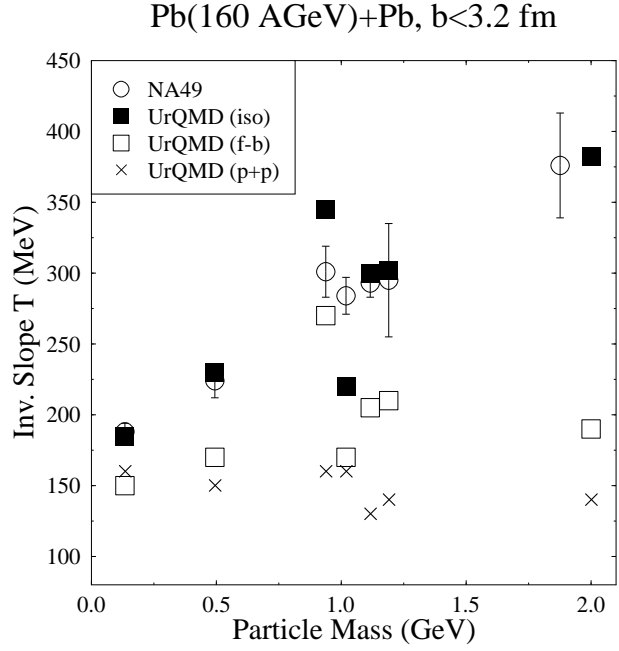,width=10cm}}
\vskip 2mm
\vspace{-0.5cm}
\caption{Midrapidity ($|y|<0.5$) inverse slope parameter T as a function of particle mass
in central Pb(160GeV)+Pb reactions (compared to p(160GeV)+p).\label{slope}}
\end{figure}

The observed apparent ''temperatures'' (inverse slopes) at midrapidity ($|y|<0.5$) 
depend strongly on the particle mass $m$ as depicted in Fig. \ref{slope}. 
The calculated proton-proton collisions (crosses) show about the same freeze-out
''temperature'' $T\approx m_{\pi}$ for all particles from $m_{\pi}$ to $m=2$~GeV.
In Pb(160 AGeV)+Pb, the ''$T$''-values increase -in contrast to the
data- only weakly above the p+p values, if the forward backward scenario
(MB) is applied. On the other hand, isotropic angular distributions for intermediate energy 
meson-baryon collisions (in line with the resonance picture)
yield a linearly increasing apparent ''temperature'' with the particle mass
for Pb(160 AGeV)+Pb, in agreement with the preliminary NA49 data.

We conclude that strong transverse flow can occur in massive
reactions at the SPS. However, the hadron mass dependence of the apparent
temperature is sensitive to the detailed modelling of the meson-baryon
rescattering process above the resonance region ($\sqrt{s_{\rm
MB}}>2$~GeV) and below the high energy 
domain. Two distinct pictures of meson baryon scattering with
$\sqrt{s_{\rm MB}}>2$~GeV have been 
confronted: the high energy limit of longitudinal fragmenting color
strings was extrapolated downward in energy, versus the concept of resonance 
formation of meson baryon scattering with isotropic decay.
Both models describe the rapidity dependence of the transverse momentum
of produced particles equally well, while the isotropic scattering prescription 
predict the apparent proton temperatures at midrapidity about 30\%
larger than preliminary data. 
However, strong increase of the transverse slope does {\em only} build up in
the high energy 'resonance
model'.

\section*{acknowledgements}
This work is supported by the BMBF, GSI, DFG and Graduiertenkolleg 'Schwerionenphysik'. 
M. Bleicher and S. Soff acknowledge support by the Josef Buchmann Foundation.

\newpage

\newpage

\end{document}